# Photon counting statistics using a digital oscilloscope

M. L. Martínez Ricci,[1,2] J. Mazzaferri,[1] A. V. Bragas,[1,2] and O. E. Martínez[1,2]

[1]Departamento de Física, Facultad de Ciencias Exactas y Naturales, Universidad de Buenos Aires, Argentina

[2]Consejo Nacional de Investigaciones Científicas y Técnicas, CONICET, Buenos Aires, Argentina

We present a photon counting experiment designed for an undergraduate physics laboratory. The statistics of the number of photons of a pseudo-thermal light source is studied in two limiting cases: much longer and much shorter than the coherence time, giving Poisson and Bose-Einstein distributions, respectively. The experiment can be done in a reasonable time using a digital oscilloscope without the need of counting boards. The use of the oscilloscope has the advantage of allowing the storage of the data for further processing. The stochastic nature of the detection phenomena adds additional value because students are forced to do data processing and analysis.

## I. INTRODUCTION

In a typical photon counting experiment, a photomultiplier tube (PMT) is used to convert light into electrons, which are amplified into electrical pulses and sent to counting electronics. In a very dim illumination experiment, each pulse is assigned to a single photon reaching the photocathode and is recorded as a count. By obtaining a large number of counts, the statistics of the number of photons per unit time can be recorded, giving valuable information about the nature of the light source.[1] For example, a classical monochromatic light source with a perfect detector would yield the same count at equal time intervals (in a semiclassical approach the counts arise from the quantum nature of the electrons). A fluctuating field such as that generated by a thermal source would yield different counts each time the experiment is repeated, following a characteristic statistical behavior that is the signature of the random behavior of thermal sources.

It is commonly believed that a photon counting experiment needs to be done with sophisticated counting electronics, including a fast amplifier and a counting board. This requirement holds for a research quality experiment. We demonstrate in this paper that the same results can be achieved with modest equipment. Moreover, additional analysis of the data adds new experimental evidence of the statistics and contributes to the learning process. The counting electronics is replaced by a digital oscilloscope, which records the time traces from the PMT at a given time window. The data for each window are acquired through the serial port of a personal computer. We show that computing the probability of counting *n* photons vs. the time windows gives a signature that helps to determine which statistics is appropriate for describing the light source. Once the information is stored in the computer, it can be analyzed in different ways, allowing a deeper understanding of statistics, correlation, and data processing. One of the advantages of measuring the time traces of the photon detection experiment is the possibility of measuring the statistical distribution of the arrival time of the photons. This measurement brings additional information that is useful for distinguishing between light sources with Bose-Einstein and Poisson statistics.

In Sec. II the theory needed to understand the analysis of counting statistics is given. The experiment is described in Sec. III and the results of our experiment are presented in Sec. IV. The general idea is to study the photon counting statistics of a pseudo-thermal light source in two limiting regimes, longer and shorter than the coherence time, which is, to some extent, controlled by the experimenter. The aim of the project is to illustrate the stochastic nature of photodetection and the statistics of thermal light in very different regimes.

## II. PHOTON COUNTING STATISTICS

A photodetector is used to gather information on the number of photons arriving within a chosen time interval T. For a light source in thermal equilibrium the emission is not constant in time but is characterized by fluctuations. A repeated experiment at equal time intervals will not always give the same reading, thus showing the statistical nature of the light source. A simple derivation of the expected photon counting distribution at an introductory level can be found in Ref. 2. A similar derivation is presented in the introductory chapter of Ref. 3. The electromagnetic field oscillating at frequency ω has energy at values

$$E_n = (n+\tfrac{1}{2})\hbar\omega, \tag{1}$$

where $n$ is the number of photons or quanta of electromagnetic radiation, and $h$ represents Planck's constant. At thermal equilibrium at temperature $\theta$ (we use the Greek letter $\theta$ to distinguish it from the time interval $T$) the probability of finding the mode excited at energy $E_n$ is given by the Boltzmann factor:

$$P(n) = \frac{e^{-E_n/k_B\theta}}{\sum_n e^{-E_n/k_B\theta}}. \tag{2}$$

Note that the zero point energy $\hbar\omega/2$ from Eq. (1) cancels in Eq. (2). The average number of photons $<n>$ can be obtained from Eq. (2) using Eq. (1):

$$\langle n \rangle = \sum_n n P(n) = \frac{\sum_n n(e^{-\hbar\omega/k_B\theta})^n}{\sum_n (e^{-\hbar\omega/k_B\theta})^n} = \frac{1}{e^{\hbar\omega/k_B\theta}-1}, \tag{3}$$

where Eq. (3) gives the average number of photons in a cavity in thermal equilibrium and is known as Planck's thermal excitation function. If we multiply Eq. (3) by the energy per photon and the density of modes in a cavity, we obtain the Planck blackbody radiation spectrum.[2,3,4,5] From Eqs. (2) and (3) the probability $P(n)$ can be expressed in terms of the average number $<n>$:

$$P(n) = \frac{\langle n \rangle^n}{(1+\langle n \rangle)^{1+n}}. \tag{4}$$

Equation (4) indicates that repeated experiments will yield a set of $n$ values according to P(n).

In practice we measure the number of photons in a time interval $T$, hence sampling a volume of length $L = cT$, because photons farther away will not reach the detector in this interval. Repeating the experiment consists in taking the same interval $T$ starting at different times *t*, assuming that the light source is stationary. The fluctuations in the measurement appear as fluctuations in the power of the light source.

The results we have discussed assume that a single mode of the electromagnetic field is measured. This assumption requires that the light can be considered monochromatic during the measurement interval; that is, the time *T* is much smaller than the coherence time of the source $T_c$,

$$T << T_c. \qquad (5)$$

The coherence time is the inverse of the bandwidth and can be determined by observing the characteristic fluctuation time of the signal, either directly if the detector is fast enough, or else by observing the contrast of an interference pattern as a function of the delay. Because the experimental apparatus requires time intervals of several microseconds, extremely monochromatic light sources are necessary to satisfy Eq. (5). A single longitudinal mode He-Ne laser is therefore used and the thermal character artificially imposed by introducing random fluctuations as described in Sec. III. If we substitute the time interval explicitly in Eq. (4), the probability of detecting *n* photons in the time *T* for a single mode and thermal light becomes

$$P_n(T) = \frac{\langle n \rangle^n}{(1+\langle n \rangle)^{1+n}} \quad (T << T_c), \qquad (6)$$

where $< n >$ is a function of *T*. From the semiclassical point of view, the light intensity fluctuates randomly with the typical duration of the fluctuations of the order of $T_c$, and the field is sampled at time intervals *T* much smaller that the duration of the fluctuations. During the time *T* the power can be considered constant, but between one measurement and the other, random fluctuations occur. Photons detected within a time *T* are correlated, which means they are not statistically independent. That is, if a photon is detected within the interval *Δt* inside *T*, it is more likely that another photon will be detected in nearby intervals *Δt* within *T*.

In the absence of fluctuations the detection of photons should not be correlated; that is, the fact that a photon has been detected in a time interval *Δt* should not influence the probability of detecting a photon at a nearby $\Delta t$. In this limit the probability of detecting a photon in an interval $\Delta t$ should be proportional to $\Delta t$ and is independent of any other event occurring at other time intervals. In this case the probability of detecting $n$ photons in a time interval $T$ should follow the Poisson distribution[6,7]

$$P_n(T) = \frac{\langle n \rangle^n}{n!} e^{-\langle n \rangle}. \tag{7}$$

Uncorrelated photons yield a Poisson distribution, but they also have other distinct statistical signatures. One clear signature can be seen in the arrival time probability $P_{at}(T)$, which is the probability of measuring two consecutive photocounts with a time delay $T$. This probability can be expressed as the product of two probabilities, the probability of counting *0* photons within the interval *[t,t + T]* multiplied by the probability of counting *1* photon at time *t + T*. After a statistical average over the fluctuations of the intensity, $P_{at}(T)$ can be expressed as

$$P_{at}(T) = \langle P_0(t,T) p(t+T) \rangle. \tag{8}$$

For uncorrelated events, $p(t)$ takes a constant value, and Eq. (8) can be expressed as

$$P_{at}(T) = \zeta\, P_0(T), \tag{9}$$

where $\zeta$ is a constant. Otherwise, $P_{at}(T)$ is not proportional to $P_0(T)$.

When using thermal light the photon correlations will be lost for times much longer than $T_c$. This phenomenon can be understood by noting that if the time interval $T$ is very large, many fluctuations take place, and hence we are measuring an average value of the fluctuations and not the fluctuation itself. The longer the interval, the closer the measured value approaches the mean value. As a consequence, the measured statistics approach the uncorrelated Poisson distribution. Another way to look at the problem is to notice that if $T \gg T_c$, the light cannot be regarded as monochromatic, and hence many modes are needed to describe it. The photons within one mode are correlated by the Bose-Einstein distribution, but photons from different modes are uncorrelated. Hence when counting photons within a large time interval, photons from different modes are detected and the correlations are lost.

In a more formal description of the two limiting cases the following discussion can be used (for more detail see Ref. 3). In the semiclassical theory of optical detection, the electromagnetic field is treated classically, and the PMT converts a classical continuous intensity $\bar{I}$ into a succession of discrete counts. With the assumption that the probability $p(t)$ per unit time of having a single count at time $t$ is proportional to the intensity $\bar{I}(t)$, the Mandel formula can be obtained:[3,8,9]

$$P_n(T) = \left\langle \frac{[\xi \bar{I}(t,T)T]^n}{n!} \exp[-\xi \bar{I}(t,T)T] \right\rangle, \qquad (10)$$

where $\xi$ is the efficiency of the detector; the distribution is obtained as a statistical average over the fluctuations of the intensity $\bar{I}(t,T)$. It is difficult to find a general expression for the statistical average of a time dependent function, but for the two cases we have discussed the expressions we have given can be readily derived.[3]

Thus far, we have obtained the behavior of a thermal source in two extreme cases. However, it is almost impossible experimentally to work with real thermal light sources under the condition $T<<T_c$, because the coherence time of real thermal light sources is much less than $10^{-8}$ s. For this reason we built a pseudo-thermal light source, in which the coherence time $T_c$ can be chosen to satisfy the conditions $T<<T_c$ and $T>>T_c$. The experimental set up is described in Sec. III.

## III. EXPERIMENTAL

A stable high voltage source at -1200 V feeds the PMT (Hamamatsu 1P28). The current signal built in the PMT passes through a load resistor $R_L$ and the voltage drop is recorded by the oscilloscope. The digital oscilloscope Tektronix TDS 360 (200 MHz bandwidth, and 1000 acquisition points per screen) is set at a given window time $T$, and each window is acquired through the RS232 port of a personal computer. A program provided by Tektronix was slightly modified to continuously acquire the data for each time window. Simple codes were written to extract the number of peaks, peak height, etc. If we used faster ports, the data could be downloaded more efficiently, allowing shorter experiments, but even with our slow port enough statistics can be gathered in a reasonable lab time.

A pseudo-thermal light source is generated using the experimental setup shown in Fig. 1, following the method of Ref. 8. The coherent light of a He-Ne laser is passed through a ground-acrylic disk, which can rotate at a selectable speed. A short focus lens *L* makes the beam converge on one point of the disk to help produce a speckle pattern that diverges away from it. This speckle pattern can be observed when the disk is still. When the disk moves, the spatial coherence of the pattern is broken at a fixed observation point. Therefore, by selecting the speed of rotation of the disk, the coherence time $T_c$ of the pseudo-thermal light can be chosen.[1]

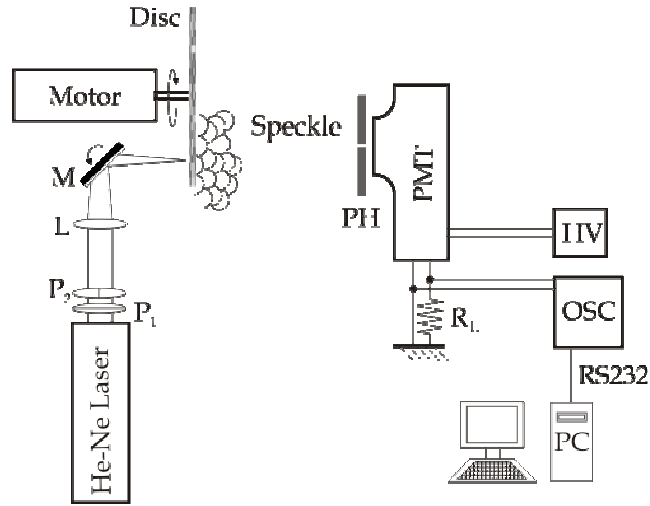

**Figure 1.** Experimental setup used for the photon counting experiment of a pseudo-thermal light source.

Two polarizers P1 and P2, with their transmission axes orthogonally oriented, are placed after the He-Ne laser. These polarizers reduce the intensity of the beam, which is an important condition of the experiment. The pinhole PH, which is attached to the PMT, not only helps to keep the counts low enough to be in a single photon counting condition, but also reduces undesirable counts from other sources. The size of the pinhole was chosen to be smaller than the speckle grain characteristic size so that the intensity fluctuations could eventually reach the zero. A mirror M placed just before the disk is used to direct the desired portion of the speckle pattern to the PMT. A sealed box is used to protect the PMT from residual ambient light, and the box has only a small hole to allow the entrance of the desired light. Selecting a region smaller than a speckle grain is equivalent to selecting a time window smaller than $T_c$, which is also analogous to select a single transverse spatial mode, needed to recover the thermal fluctuations as discussed in Sec. II. If the pinhole is enlarged so as to average over many grains, the fluctuations are smoothed out and the Poisson distribution is obtained.

## IV. RESULTS AND DISCUSSION

**A. Characterization of the intensity fluctuations**

The goal of the experiment is to measure the statistics of the number of photons per unit time $T$ of a pseudo-thermal ligh. As discussed in Sec. II, Bose-Einstein statistics should be observed for $T << T_c$ and Poisson statistics for $T >> T_c$. To measure the light intensity fluctuations of the pseudo-thermal light source and, hence, $T_c$, we used the same setup shown in Fig. 1, but with the PMT used as an intensity photodetector. For this purpose, the PMT output signal is connected directly to the oscilloscope, so that the oscilloscope high input impedance acts as the load resistor giving a long integration time. Figure 2 is a typical time trace showing the intensity fluctuations of the pseudo-thermal light, for which the angular frequency of the disk is set at $\omega = 25$ mHz. For this experiment we acquired 103 windows of $T = 1$ s, and the statistics of the intensities were recorded. As can be seen from Fig. 3, the intensities behave as predicted for thermal light: a negative exponential with the highest probability at zero intensity.[3] The characteristic time $T_c$ is the inverse of the characteristic frequency of the intensity fluctuations. Each time window has associated with it a characteristic frequency that is obtained by averaging its Fourier transform over the frequencies. The histogram of the characteristic frequencies among the time windows is a Gaussian-shaped distribution. The characteristic frequency of the intensity fluctuations is defined as the peak value of the histogram. This experiment is useful for determining the $T_c$ range that can be attained with a given experimental setup. For our setup we obtained $T_c = 4 \times 10^{-4}/\omega$, giving a range of 7 μs < $T_c$ < 17 ms for realistic disc rotation speeds.

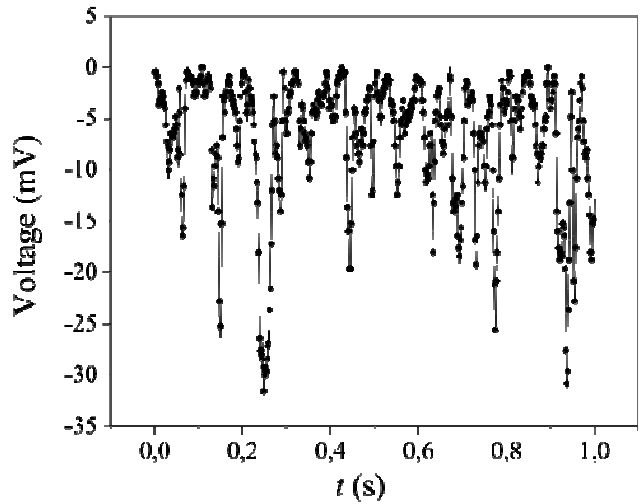

**Figure 2.** Typical time trace picturing the intensity fluctuations (expressed in mV) of the pseudo-thermal light source; Tc ≅ 20 ms.

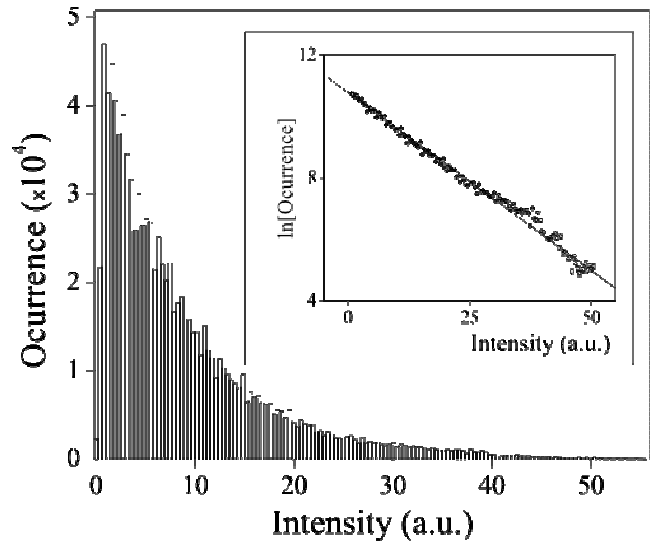

**Figure 3.** Statistics of the intensity values of 103 registered events. The linear behavior of the natural logarithm shown in the inset demonstrates that the light source is thermal.

## B. Counting photons

The PMT gain, which depends on the bias voltage, defines the pulse peak voltage, which can be estimated from the datasheet of the PMT. However, the stochastic process generating secondary electrons inside the PMT gives a Poisson distribution for the peak voltages. A typical screen shot of the oscilloscope in photon counting mode, with $R_L = 50\ \Omega$ and $T = 2\ \mu s$, is shown in Fig. 4(a). Each peak in the screen corresponds to either a photon reaching the photocathode or to a spurious noise peak. To choose a threshold above which a pulse will be counted as a photon-count, we plot the peak height histogram as shown in Fig. 4(b). It is clear that the peak pulse distribution is separated from the noise contribution which is notoriously higher than the former for low voltages.

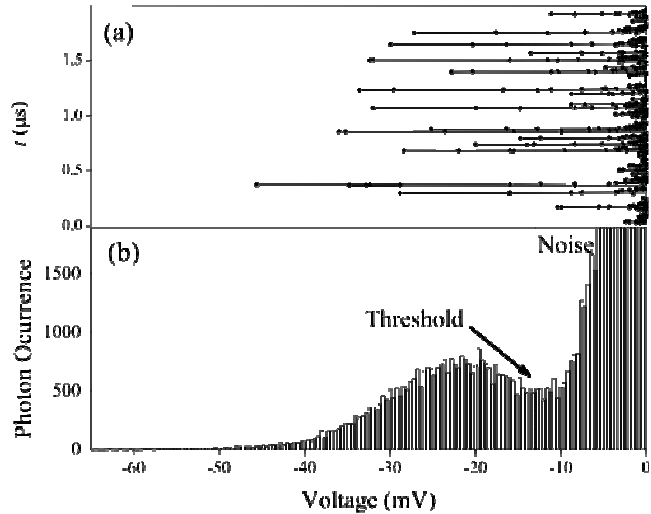

**Figure 4.** (a) Typical screen shot of the oscilloscope in photon counting mode, with $R_L = 50\ \Omega$ and $T = 2\ \mu s$. (b) Statistics of the height of the photocounts. The noise contribution is mostly separated from the photocounts peaks, allowing the definition of an appropriate threshold.

## C. Bose-Einstein and Poisson statistics

We performed two experiments to explore the two statistical behaviors: the Bose-Einstein and the Poisson regimes. We used the same pseudo-thermal light source and chose different experimental parameters for each experiment as indicated in Table I. The highest $T_c$ available was used for the Bose-Einstein and the smallest $T_c$ for the Poisson experiments. The time window $T$ for each experiment was chosen to satisfy the corresponding conditions. Due to the finite size of the oscilloscope register, the time window choice determines the time resolution of the experiment. To sample the electrical pulses corresponding to one photon with at least two points, the width of the pulse has to be adjusted according to the time windows. The shape and width of the pulses are defined by the building time of the PMT (~ 5 ns), the load resistor, $R_L$, and the load and cable capacitance. We selected the $R_L$ values in Table I to control the width of the pulses in each experiment. Figure 5 depicts the typical shape of the pulses in the two experiments. In both experiments the beam intensity was controlled to maintain a low enough average number of photons per unit time to avoid the superposition of pulses.

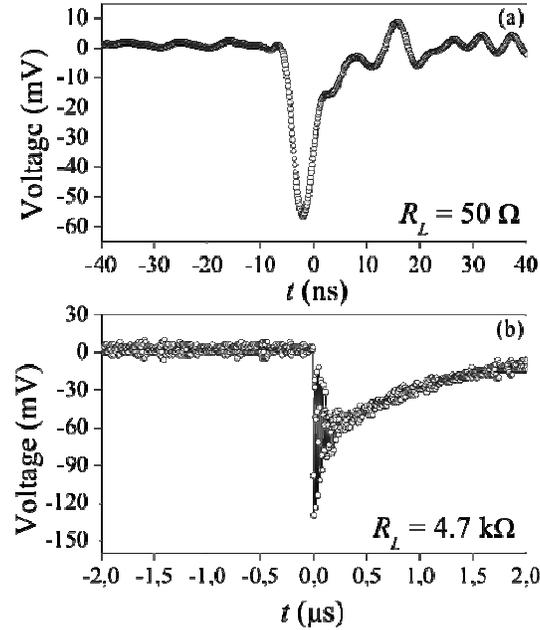

**Fig. 5.** A unique photocount. (a) $R_L$ = 50 Ω for the Bose-Einstein experiment; (b) $R_L$ = 4.7 kΩ for the Poisson experiment.

|  | Bose-Einstein | Poisson |
|---|---|---|
| ω | 25 mHz | 66 Hz |
| $T_c$ | 17.54 ms | 6.61 μs |
| T | 2 μs | 1 ms |
| $R_L$ | 50 Ω | 4.7 kΩ |
| Photon pulse width | 20 ns | 2 μs |

**Table I.** Parameters used to set appropriate conditions for the two limiting cases.

Figure 6 shows the histogram of the number of photons recorded in the two experiments, respectively. The average number of photons <n> in each case can be determined from Fig. 6 and corresponds to 1.6 photon/μs in the Bose-Einstein and 18.4 photons/ms in the Poisson experiment. The error bars were estimated by taking into account the dark noise for this PMT (measured dark noise: 800 photons/s), the arrival of spurious light, and the statistical fluctuations.

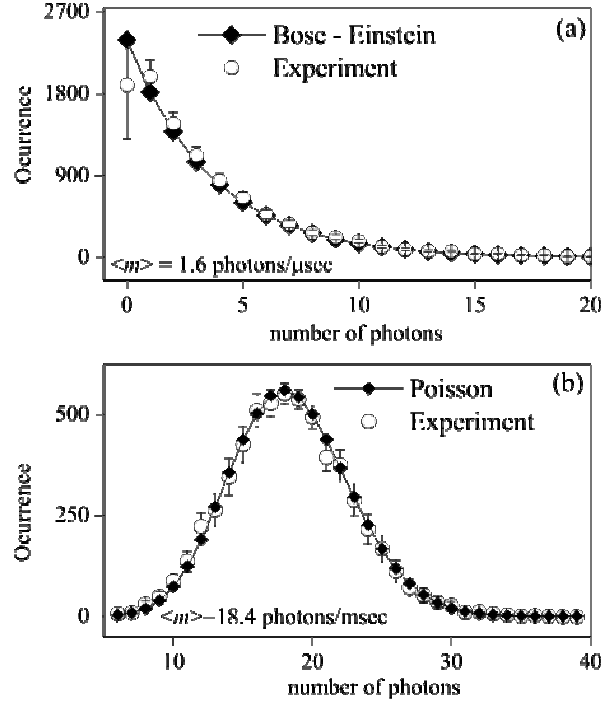

**Figure 6.** (a) Bose-Einstein distribution. (b) Poisson distribution.

As can be seen from Fig. 6, the experiment accurately confirms the predictions of Eqs. (6) and (7). The initial characterization of the quality of the fit is made by applying the $\chi^2$ test, giving a confidence of 90% for the Bose-Einstein experiment and 70% for the Poisson case. The number of time windows considered was 10 000 for Bose-Einstein and 6000 for Poisson. The reason for the lower confidence in the latter case is not fully understood. One reason could be that, due to the experimental constrains, we could not fully achieve the condition $T >> T_c$. Nevertheless, a subestimation of the error bars would produce worse confidence. In any case, the fitting is good enough to decide for one or the other statistics. By measuring the time traces of the photon detection process with the oscilloscope, we have additional information that gives further support to the photon counting measurements.

### D. Measurements of $P_0$, $P_1$, and $P_2$

In Sec. IVC we showed convincing evidence that we were detecting both regimes of pseudo-thermal light, well above and well below $T_c$. However, due to similarities of the two statistics the predicted histograms look similar for a very low average number of photons. We now show that by analyzing the stored data in a different manner better discrimination can be obtained, thus gaining confidence in the models. We analyze the probability $P_0$, $P_1$, and $P_2$ of counting 0, 1, and 2 photons, respectively, as a function of the time window. The time traces acquired in two experiments were divided into smaller time subwindows ($\tau$) and the number of peaks for each $\tau$ was counted.

To compute the probabilities $P_0$, $P_1$, and $P_2$ for the two conditions, we use the following expressions, derived from Eqs. (7) and (6) respectively:

$$P_n(\tau) = \frac{[\frac{\langle n \rangle_T}{T}\tau]^n}{n!} \exp\left(-\frac{\langle n \rangle_T}{T}\tau\right) \quad (n = 0, 1, 2,\ldots) \text{ (Poisson)} \quad (11)$$

$$P_n(\tau) = \frac{(\frac{\langle n \rangle_T}{T}\tau)^n}{(1+\frac{\langle n \rangle_T}{T}\tau)^{1+n}} \quad (n = (0, 1, 2,\ldots) \text{ (Bose-Einstein)}, \quad (12)$$

where $\langle n \rangle_T$ is the average number of photons in the window time $T$. Figure 7 shows the occurrences of 0, 1, and 2 photons as a function of the time window $\tau$, in the range of 20 ns to 1 $\mu$s, for the Bose-Einstein data. The theoretical curves $P_0(\tau)$, $P_1(\tau)$, and $P_2(\tau)$ are also shown. For a given (and even rather low) average number of counts the two statistics behave very differently and the data can be clearly associated with one of the two. From Fig. 7 the results are conclusive in that the experimental points overlap only the theoretical prediction corresponding to Bose-Einstein statistics. The same analysis was done with the Poisson data shown in Fig. 8, with the time window $\tau$ up to 400 $\mu$s. The results are again conclusive, indicating that the light source analyzed in this experiment follows Poisson statistics.

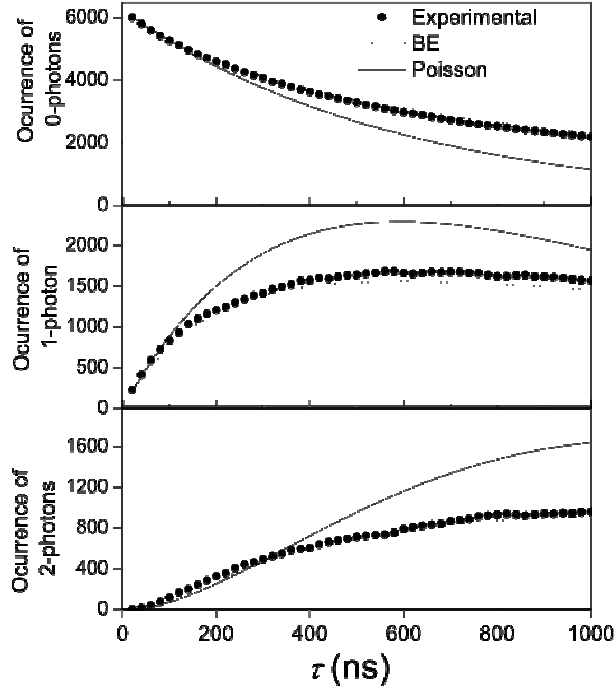

**Figure 7.** Frequency of n = 0, 1, and 2 photons versus the time window size τ for the data of the Bose-Einstein experiment I. The theoretical prediction of Bose-Einstein statistics fits the experiment data but not the theoretical prediction of Poisson statistics.

**IVE. Measurements of arrival time**

Using Eq. (9), an additional test that can distinguish between the two statistics can be performed. The main idea is to check if $P_0(T)$ (the probability of detecting 0 photons in a $T$ window) and $P_{at}(T)$ (the arrival time probability) are proportional. In such a case the source follows Poisson statistics, otherwise the source is described by some other statistics.

Figure 9(a) shows the occurrences of arrival time intervals for the Bose-Einstein data, normalized to the total number of computed intervals. Also shown is the occurrence of 0 photons, normalized to the number of computed time windows with 0 photons. It is evident from Fig. 9(a) that the two curves are not proportional, and therefore the corresponding experiment does not follow Poisson statistics as expected. Figure 9(b) shows that the curves are proportional to each other, verifying the intrinsic Poisson statistics.

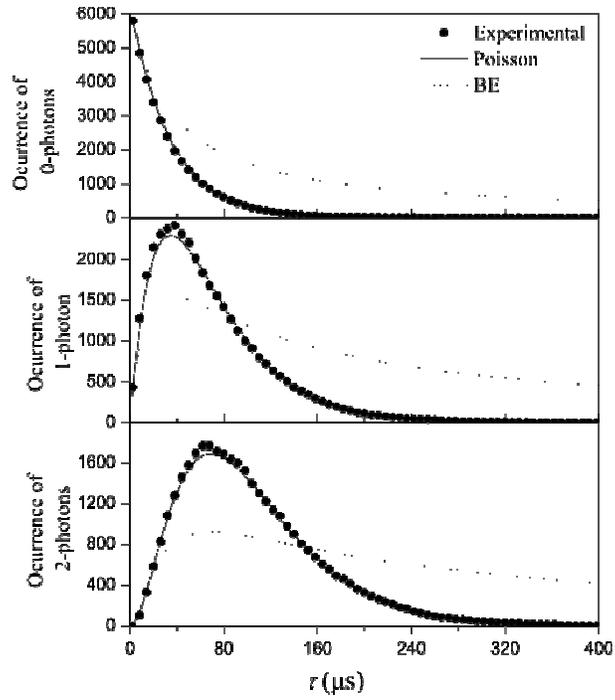

**Figure 8.** Frequency of n = 0, 1, and 2 photons versus τ for the data of the Poisson experiment. The theoretical prediction of Poisson statistics fits the experiment data but not the theoretical prediction of Bose-Einstein statistics.

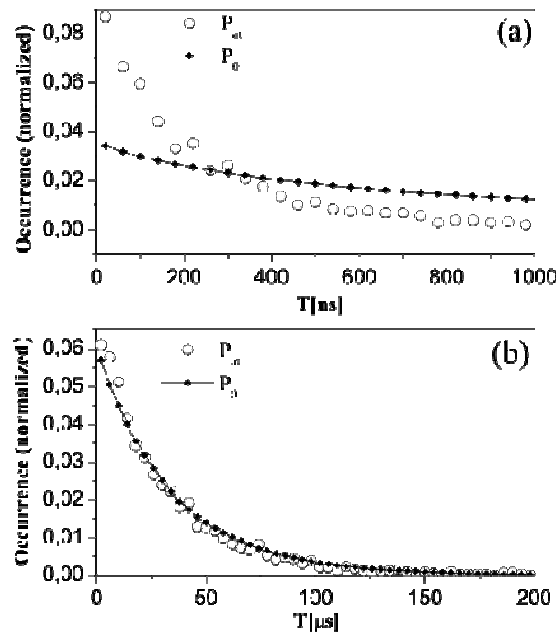

**Fig. 9.** Comparison between $P_0$ and $P_{at}$ for (a) the Bose-Einstein experiment, (b) The proportionality present in (b) reaffirms the Poisson behavior.

## V. CONCLUSIONS

A pseudo-thermal light source was studied in two photon counting experiments. The parameters in each experiment were controlled to reach two limiting situations, and Bose-Einstein and Poisson statistics were clearly observed. Modest undergraduate lab equipment was employed to obtain the time traces of the photon detection process. The results were first studied with traditional photon counting procedures and the two statistics could be observed. Because the time traces of the experiment were recorded, additional information on the process could be computed. The probabilities of 0, 1, and 2 photons as a function of the time windows were measured. Also, the arrival time of photons was studied and compared with the probability of 0 photons. The two additional tests gave conclusive confirmation of the results obtained with the traditional methodology.

## ACKNOWLEDGMENTS

This work was done as a lab project in a quantum optics course given at the Physics Department of the School of Science of the University of Buenos Aires. We thank the Physics Department for its support, and lab technicians Alejandro Greco and Fernando Monticelli for their help. We also thank Professor Claudio Iemmi for supplying some of the material. M.L. Martínez Ricci and J. Mazzaferri are fellows of CONICET and Universidad de Buenos Aires, respectively.